\documentclass[aps,prl,twocolumn,superscriptaddress,showpacs]{revtex4}

\usepackage{amsmath}
\usepackage{graphics}
\usepackage{natbib}

\begin{document}

\title{Bistability of Anderson localized states in nonlinear random media}

\author{Ilya V. Shadrivov}

\affiliation{Nonlinear Physics Center, Research School of Physics and
Engineering, Australian National University, Canberra ACT 0200, Australia}

\author{Konstantin Y. Bliokh}

\affiliation{Nonlinear Physics Center, Research School of Physics and Engineering,
Australian National University, Canberra ACT 0200, Australia}
\affiliation{Applied Optics
Group, School of Physics, National University of Ireland, Galway, Galway, Ireland}

\author{Yuri S. Kivshar}

\affiliation{Nonlinear Physics Center, Research School of Physics and
Engineering, Australian National University, Canberra ACT 0200, Australia}

\author{Yuri P. Bliokh}

\affiliation{Department of Physics, Technion-Israel Institute of Technology, Haifa 32000,
Israel}

\author{Valentin Freilikher}

\affiliation{Department of Physics, Bar-Ilan University, Ramat-Gan 52900, Israel}

\begin{abstract}
We study wave transmission through one-dimensional random nonlinear structures and
predict a novel effect resulting from an interplay of nonlinearity and disorder. We
reveal that, while weak nonlinearity does not change the typical exponentially small
transmission in the regime of the Anderson localization, it affects dramatically the
disorder-induced localized states excited inside the medium leading to {\em bistable} and
{\em nonreciprocal} resonant transmission. Our numerical modelling shows an excellent
agreement with theoretical predictions based on the concept of a high-Q resonator
associated with each localized state. This offers a new way for all-optical light control
employing statistically-homogeneous random media without regular cavities.
\end{abstract}

 \maketitle

The localization of waves in disordered media, also known as Anderson localization, is a
universal phenomenon predicted and observed in a variety of classical and quantum wave
systems~\cite{Localization}. Recent renewed interest to this phenomenon is driven by a
series of experimental demonstrations in optics~\cite{optics} and Bose-Einstein
condensates~\cite{BEC}. One of the important issues risen in these studies is that the
disordered systems can be inherently nonlinear, so that an intriguing interplay of
nonlinearity and disorder could be studied experimentally.

Nonlinear interaction between the propagating waves and disorder can significantly change
the interference effects, thus fundamentally affecting
localization~\cite{nonlin-loc1,nonlin-loc2}. However, most of the studies of the
localization in random nonlinear media deal with the ensemble-averaged characteristics of
the field, such as the mean field and intensity, correlation functions, etc. These
quantities describe the averaged, typical behavior of the field, but they do not contain
information about \textit{individual localized modes} (resonances), which exist in the
localized regime in each realization of the random sample
\cite{Frisch-Azbel,we-JOSA,we-PRL-1,Topolancik}. These modes are randomly located in both
real space and frequency domain and are associated with the exponential concentration of
energy and resonant tunnelling. In contrast to regular resonant cavities, the Anderson
modes occur in a \textit{statistically-homogeneous} media because of the interference of
the multiply scattered random fields. Although the disorder-induced resonances in linear
random samples have been the subject of studies for decades, the resonant properties of
nonlinear disordered media have not been explored so far.

In this Letter we study the effect of nonlinearity on the Anderson localized states in a
one-dimensional random medium. As a result of interplay of nonlinearity and disorder, 
bistability and nonreciprocity appear upon resonant wave tunnelling and excitation of
disorder-induced localized modes in a manner similar to that for regular cavity modes. At
the same time, weak nonlinearity has practically no effect on the averaged localization
background.


First, we consider a stationary problem of the transmission of a monochromatic wave
through a one-dimensional random medium with Kerr nonlinearity. The problem can be reduced to the equation
\begin{equation}
\label{oscillator}
\frac{d^2\psi}{dx^2}+k^{2}\left[n^2 -\chi|\psi|^2 \right]\psi=0~,
\end{equation}
where $\psi$ is wave field, $x$ is coordinate, $k$ is wave number in the vacuum, $n=n(x)$
is the refractive index of the medium, and $-\chi $ is the Kerr coefficient.

In the linear regime, $\chi|\psi|^2 = 0$, the multiple scattering of the wave on the
random inhomogeneity $n^2(x)$ brings about Anderson localization. The main signature of
the localization is an exponential decay of the wave intensity, $I=|\psi|^2$, deep into
the sample and, thus, an exponentially small transmission \cite{Localization,Berry}:
$I_{\mathrm{out}}^{\mathrm{(typ)}}\sim I_{\mathrm{in}}\exp(-2L/l_{\mathrm{loc}})\ll 1$.
Here $L$ is the length of the sample and $l_{\mathrm{loc}}$ is the localization length
which is the only spatial scale of Anderson localization. Along with the typical wave
transmission, there is an anomalous, resonant transmission, which accompanies excitation
of the Anderson localized states inside the sample and occurs at random resonant wave
numbers $k=k_{\mathrm{res}0}$ \cite{Frisch-Azbel,we-JOSA,we-PRL-1,Topolancik}. In this
case, the intensity distribution in the sample is characterized by an exponentially
localized high-intensity peak inside the sample, $I_{\mathrm{peak}} \gg I_{\mathrm{in}}$,
and a transmittance much higher than the typical one:
$I_{\mathrm{out}}^{\mathrm{(res)}}\gg I_{\mathrm{out}}^{\mathrm{(typ)}}$ (see Fig.~1).

\begin{figure}[t]
\centering \scalebox{0.48}{\includegraphics{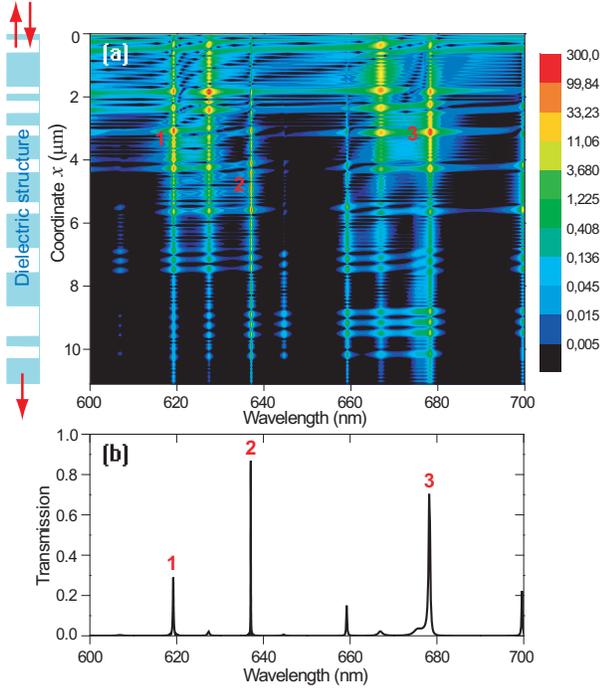}} \caption{(Color online.)
Excitation of localized modes in the linear regime: (a) normalized intensity of the
field, $I/I_{\rm in}$ (log scale), vs. the wavelength and position in the sample and
(b) the transmission spectrum.}
\label{Fig1}
\end{figure}

Excitation of each localized mode inside the random sample can be associated with an
effective resonator cavity located in the area of field localization and characterized by high quality factor $Q\gg 1$ \cite{we-rev}. According to this model, the transmittance spectrum $T(k)$ in the vicinity of a resonant wavelength,
$|k-k_{\mathrm{res}0}|\ll k_{\mathrm{res}0}$ is given by the Lorentzian dependence \cite{we-JOSA,Genack}:
\begin{equation}
\label{Lorentz_transmission} 
T(k)=\frac{T_{\mathrm{res}}}{1+\left[2Q\left({k}/{k_{\mathrm{res}0}}
-1\right)\right]^2}~,
\end{equation}
where $T_{\mathrm{res}}=T(k_{\mathrm{res}0})$.

Let us account now for the effect of weak nonlinearity: $|\chi\psi^2| \ll 1$. The nonlinearity becomes noticeable, first of all, at the resonant intensity peaks $I_{\mathrm{peak}}$ inside the sample. It is physically
clear that the Kerr term in Eq.~(\ref{oscillator}) changes the effective refractive index of the medium
leading to the \textit{intensity-dependent shift} of the resonant wave number:
$k_{\mathrm{res}0} \rightarrow k_{\mathrm{res}} (I_{\mathrm{peak}})$. Since the values of
$I_{\mathrm{peak}}$ and $I_{\mathrm{out}}$ are unambiguously connected, the resonant wave
number is a function of the output intensity, which in the case of weak nonlinearity
takes the form:
\begin{equation}
\label{res_frequency}
k_{\mathrm{res}} (I_{\mathrm{out}})\simeq k_{\mathrm{res}0} + \left.\frac{%
d k_{\mathrm{res}}}{d I_{\mathrm{out}}}\right|_{I_{\mathrm{out}}=0}I_{%
\mathrm{out}}~.
\end{equation}
Substitution Eq.~(\ref{res_frequency}) into Eq.~(\ref{Lorentz_transmission}) yields
\begin{equation}
\label{nonlinear_transmission} 
T(k,I_{\mathrm{out}})\equiv
\frac{I_{\mathrm{out}}}{I_{\mathrm{in}}}=\frac{T_{\mathrm{res}}}{1+\left[A\chi
I_{\mathrm{out}}+\delta\right]^2}~.
\end{equation}
Here we introduced two dimensionless parameters $A$ and $\delta$, which characterize,
respectively, the strength of the nonlinear feedback and the detuning from the resonant
wave number:
\begin{equation}
\label{params}
A= \frac{2Q}{\chi} \left.\frac{d \ln k_{\mathrm{res}}}{d I_{\mathrm{out}}}\right|_{I_{%
\mathrm{out}}=0},~ \delta=2Q\left(1-\frac{k}{k_{\mathrm{res}0}}\right).
\end{equation}

\begin{figure}[t]
\centering \scalebox{0.42}{\includegraphics{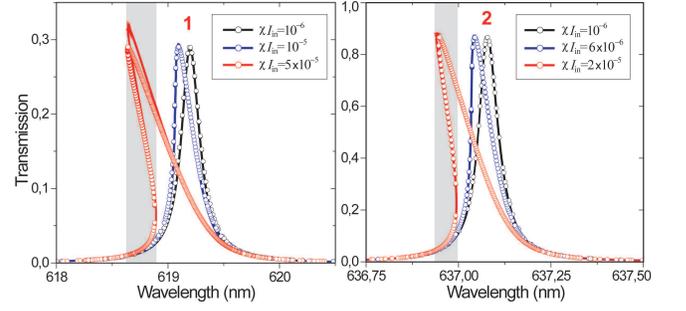}} \caption{(Color online.)
Nonlinear deformations of the transmission spectra of the resonances 1 and 2 from Fig. 1
at different intensities of the incident wave: numerical simulations of Eq.~(1)
(curves) and theoretical Eq.~(4) (symbols). Light-grey stripes indicate three-valued regions for the
high-intensity curves, where only two of them (corresponding to the lower and upper
branches) are stable.} \label{Fig2}
\end{figure}

\begin{figure}[t]
\centering \scalebox{0.42}{\includegraphics{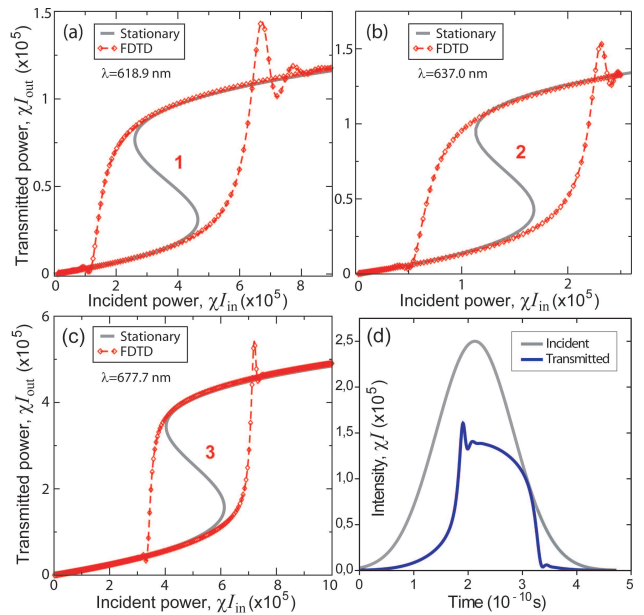}} \caption{(Color online.)
Stationary and FDTD simulations showing hysteresis loops in the output vs. input power
dependence for resonances 1,2, and 3 in Fig.~\ref{Fig1}. Panel (d) shows deformation of the transmitted Gaussian pulse corresponding to the hysteresis switching on the resonance 2.}
\label{Fig3}
\end{figure}

Equation (\ref{nonlinear_transmission}) establishes relation between the input and output wave intensities, which is cubic with respect to $I_{\mathrm{out}}$. It has a universal form typical for nonlinear resonators possessing \textit{optical bistability} \cite{Bistability}. From Eq.~(\ref{params}) it follows that in the region of parameters:
\begin{equation}
\label{eq6} A\delta < 0~,~~\delta^2 >3~,~~|\chi| I_{\rm in} >
\frac{8}{3\sqrt{3}}\frac{1}{|A|T_{\rm res}}~,
\end{equation}
the dependence $I_{\mathrm{out}}(I_{\mathrm{in}})$ is of the S-type and the stationary
transmission spectrum $T(k)$ is a three-valued function. In most cases, one of the
solutions is unstable, whereas the other two form a hysteresis loop in the
$I_{\mathrm{out}}(I_{\mathrm{in}})$ dependence (see Figs.~\ref{Fig2} and \ref{Fig3}).

It is important to emphasize two features of the equations (\ref{nonlinear_transmission}) and (\ref{params}), describing the nonlinear resonant transmission through a localized state. First, they have been derived without any approximations apart from the natural smallness of the nonlinearity and Lorentzian shape of the spectral line. Second, although the resonant transmission, the effect of nonlinearity, and bistability owe their origin to the excitation of the
Anderson localized mode \textit{inside} the sample, equations (\ref{nonlinear_transmission}) and (\ref{params}) contain only
quantities which can be found via \textit{outside} measurements. Indeed,
$T_{\mathrm{res}}$, $ k_{\mathrm{res}0}$, and $Q$ are determined from the transmission
spectrum in the linear regime, Eq.~(\ref{Lorentz_transmission}), while the derivative 
$\left. d \ln k_{\mathrm{res}}/ d I_{\mathrm{out}} \right|_{I_{\mathrm{out}}=0}$ can be retrieved from the shift of
the spectral line when the intensity is changed. This enables one to obtain the whole
dependence $I_{\mathrm{out}}(I_{\mathrm{in}},k)$ for any given resonance performing
external measurements of $T(k)$ at only two different intensities of the incident wave.

To verify theoretical predictions, we numerically model the transmission of light incident from $x=0$ through a random sample consisting of $N=19$ alternating layers with dielectric constants $n_1^2=1$ and $n_2^2=10$ and random widths uniformly distributed in the range $(0.12,1.08)\mu{\rm m}$, Fig.~1. This corresponds to $l_{\rm loc}=3.53\mu{\rm m}$.
Figure \ref{Fig2} shows nonlinear deformations of the resonant transmission spectra $T(k)$ for different values of $\chi I_{\rm in}>0$, which exhibit transitions to bistability. The analytical dependence $T(k)$ given by Eqs.~(\ref{nonlinear_transmission}) and (\ref{params}) with the parameters $T_{\rm res}$, $Q$, and $A$ found from the numerical experiments are in excellent agreement with the direct numerical solutions of Eq.~(\ref{oscillator}). In numerical simulations of stationary regime we used the standard 4-th order Runge-Kutta method. We note, that the incident field amplitude is a single-valued function of the transmitted field. Thus, we solve second-order ordinary differential equation Eq.~(\ref{oscillator}) using transmitted field as the initial condition for the equation.

The dimensionless parameters $T_{\mathrm{res}}$, and $Q$ from Eqs. (\ref{nonlinear_transmission}, \ref{params}), can also be estimated from a simple resonator model of the Anderson localized states \cite{we-JOSA,we-PRL-1,we-rev}:

\begin{equation}
\label{param_estimates} T_{\mathrm{res}}=\frac{4T_1 T_2}{\left(T_1+T_2\right)^2}~,~~Q^{-1}\sim\frac{
T_1+T_2}{4k_{{\rm res}0}l_{\rm loc}}~,
\end{equation}
where
\begin{equation}
\label{barrier_transmissions}
T_1 \sim \exp\left[-2x_{\mathrm{res}}/l_{\mathrm{loc}}\right]~,~~ T_2 \sim
\exp\left[-2(L-x_{\mathrm{res}})/l_{\mathrm{loc}}\right]~
\end{equation}
are the transmission coefficients of the two barriers that form the effective resonator, $x_{\mathrm{res}}$ is the coordinate of the center of the area of field localization, $l_{\mathrm{loc}}$ is the localization length, and $L$ is the length of the sample. 
 
Introducing a weak Kerr nonlinearity into the resonator model, one can also estimate the
nonlinear feedback parameter $A$:
\begin{equation}
\label{nonlin_coeff}
A\sim Q\,/\,T_2\,\overline{n^2}~,
\end{equation}
where $\overline{n^2}$ is the mean value of $n^2(x)$. 

It is important to note that {\em each disorder-induced resonance} is associated with its own effective cavity, so that the disordered sample can be considered as a chain of randomly located coupled resonators~\cite{Ref-exp-2}.

Equations (7)--(9) enable one to estimate the values of the parameters describing the nonlinear resonant wave tunnelling in Eqs.~(4) and (5) by knowing only the basic parameters of the localization -- the localization coordinate and the localization length. In particular, substituting Eqs.~(7)--(9) into Eq.~(6) and taking into account that the most pronounced transmission peaks correspond to the localized states with
$x\simeq L/2$ and $T_1 \sim T_2$, we estimate the incident power needed for bistability
of localized states:
\begin{equation}
\label{power} |\chi| I_{\rm in} \gtrsim \frac{\exp(-2L/l_{\rm loc})}{k_{{\rm res}0}l_{\rm
loc}}~.
\end{equation}
For the parameters used in our simulations this gives quite reasonable value
$|\chi|I_{\rm in}\gtrsim 10^{-5}$. If we increase the length of the sample, the Q-factors
of the resonances grow, and the incident power needed to observe bistability becomes
smaller.

To demonstrate temporal dynamics upon the bistable resonant tunnelling, we implemented an
explicit iterative nonlinear finite-difference time-domain (FDTD) scheme for modelling
pulse propagation through the disordered nonlinear sample. For precise modelling of the
spectra of narrow high-Q resonances, we employed fourth-order accurate algorithm, both in
space and in time~\cite{fourthorder}, as well as the Mur boundary conditions to simulate
open boundaries and total-field/scattered-field technique for exciting the incident wave.
Sufficient accuracy was achieved by creating a dense spatial mesh of 300 points per
wavelength ($dx=\lambda/300$). To assure stability of the method in nonlinear regime, we
used the time step of $dt = dx/3c$, and each simulation ran for $N=2*10^8$ time steps. To
compare the results of the FDTD simulations with the steady-state theory, we consider
transmission of long Gaussian pulses with central frequencies and amplitudes satisfying
conditions (\ref{power}), see Fig.~\ref{Fig3}(d). With an appropriate choice of the signal frequencies, we observe \textit{hysteresis loops} in the $I_{\mathrm{out}}(I_{\mathrm{in}})$ dependences which are in excellent agreement with stationary calculations, as shown in Figs.~\ref{Fig3}(a--c). Transitional oscillations typical for bistable nonlinear structure accompany jumps between two stable branches \cite{trans}, and strong reshaping of the transmitted pulse evidences switching between the two regimes of transmission, Fig.~\ref{Fig3}(d). Period of the transitional oscillations is defined by mismatch between external wave frequency and nonlinear eigenfrequency, whereas the decay rate is defined by the Q-factor. We note, that different choice of the signal frequencies near the resonance can lead to various other behaviours of output vs. input curves, with transmission either increasing, when nonlinear resonance frequency shifts towards the signal frequency, or decreasing in the opposite case. 

\begin{figure}[t]
\centering \scalebox{0.43}{\includegraphics{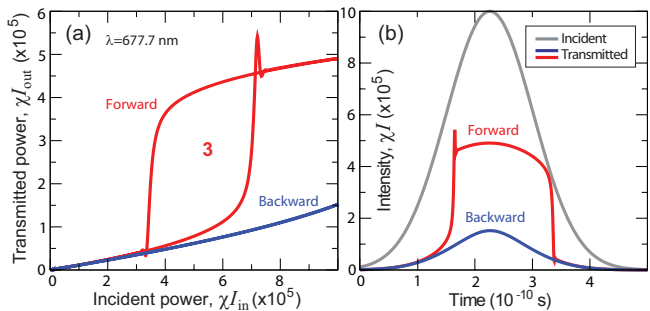}} \caption{(Color online.) (a)
Non-reciprocal transmission through the nonlinear disordered structure, showing different
output powers for identical waves incident from different directions. (b) Corresponding
shape of the incident pulse, and pulses transmitted in different directions.}
\label{Fig4}
\end{figure}

In addition to bistability, the resonant wave tunnelling through a nonlinear
disordered structure is \textit{nonreciprocical}. As is known for regular systems, {\em
nonsymmetric} and {\em nonlinear} systems may possess nonreciprocal transmission
properties, resembling the operation of a diode. An all-optical diode is a device that
allows unidirectional propagation of a signal at a given wavelength, which may become
useful for many applications~\cite{trans,diode}. A disordered structure is naturally asymmetric
in the generic case, and one may expect a nonreciprocal resonant transmission in the
nonlinear case. To demonstrate this, we modelled propagation of an electromagnetic pulse
impinging the same sample from different sides and monitor the transmission
characteristics. One case of such nonreciprocical resonant transmission is shown in
Fig.~4. We observe considerably different transmission properties in opposite
directions with the maximal intensity contrast between two directions 7.5:1. Moreover,
the threshold of bistability is also significantly different for two directions:
there is a range of incident powers, for which the wave incident from one side of the
sample is bistable, while there is no signs of bistability for the incidence from the
other side. Figure~\ref{Fig4}(b) shows the pulse reshaping for incidence from opposite
sides of the structure.

To conclude, we have studied the wave transmission through a weakly-nonlinear
statistically homogeneous one-dimensional random medium and demonstrated novel
manifestations of the interplay between nonlinearity and disorder. We have shown that
even weak nonlinearity affect dramatically the resonant transmission associated with the
excitation of the Anderson localized states leading to bistability and nonreciprocity.
Despite random character of the appearance of Anderson modes, their behavior and
evolution are rather deterministic, and, therefore, these modes can be used for efficient
control of light similar to regular cavity modes. Numerical modelling shows an excellent
agreement with theoretical analysis based on the concept of a high-Q resonator associated
with each localized state. Our results demonstrate that, unlike infinite systems, the
Anderson localization in finite samples is not destroyed by weak nonlinearity -- instead
it exhibits new intriguing features typical for resonant nonlinear systems.

The authors acknowledge a support of the Australian Research Council (Linkage International
and Discovery projects), European Commission (Marie Curie fellowship), and
the Science Foundation Ireland.



\end{document}